\documentstyle[12pt,draft,epsf]{article}
\newcommand{\resection}[1]{\setcounter{equation}{0}\section{#1}}

\def\s {\sigma}

\def\be{\begin{equation}}
\def\ee{\end{equation}}
\def\bea{\begin{eqnarray}}
\def\eea{\end{eqnarray}}
\def\beano{\begin{eqnarray*}}
\def\eeano{\end{eqnarray*}}
\def\bd{\begin{displaystyle}}
\def\ed{\end{displaystyle}}
\def\ba{\begin{array}}
\def\ea{\end{array}}

\def\nv{\vec {\bf n}}

\def\nls{nl$\s\;$}

\def\x{|_{x=0}}

\def\pd{\partial}

\begin{document}
\oddsidemargin 5mm
\setcounter{page}{0}
\newpage     
\setcounter{page}{0}
\begin{titlepage}
\begin{flushright}
CLNS 01/1750\\
\end{flushright}
\vspace{0.5cm}
\begin{center}
{\large {\bf Integrable Boundary Conditions and Reflection Matrices for 
the \\
$O(N)$ Nonlinear Sigma Model}}\\
\vspace{1.5cm}
{\bf M. Moriconi} \footnote{\tt{email:moriconi@mail.lns.cornell.edu}}\\
{\em Newman Laboratory of Nuclear Studies, Cornell University}\\
{\em Ithaca, New York 14853, USA}\\
\vspace{0.8cm}

\end{center}
\renewcommand{\thefootnote}{\arabic{footnote}}
\vspace{6mm}

\begin{abstract}
\noindent
We find new integrable boundary conditions, depending on a free
parameter $g$, for the $O(N)$ nonlinear $\sigma$ model, which are of
nondiagonal type, that is, particles can change their ``flavor''
through scattering off the boundary. These boundary conditions are
derived from a microscopic boundary lagrangian, which is used to
establish their integrability, and exhibit integrable flows between
diagonal boundary conditions investigated previously. We solve the
boundary Yang-Baxter equation, connect these solutions to the boundary
conditions, and examine the corresponding integrable flows.
\vspace{3cm}

\end{abstract}
\vspace{5mm}
\end{titlepage}

\newpage
\setcounter{footnote}{0}
\renewcommand{\thefootnote}{\arabic{footnote}}

\resection{Introduction} 

One of the main problems in boundary integrable field theories is to
find boundary conditions that preserve the integrability of a given
bulk integrable field theory, that is, integrable boundary conditions.
This has been carried out for several models
by direct inspection of higher-spin conserved charges 
\cite{GZ,BCDR,MDM,AEG}. In \cite{MDM} (see also \cite{M,MS}) it has 
been shown that models with $O(N)$ or $SO(N)$
symmetry admit a family of $N+1$ integrable boundary conditions, which
are obtained by requiring $n$ field components to satisfy Neumann and
the remaining $N-n$ to satisfy Dirichlet, $n=0,1, \ldots, N$. These
conditions are of diagonal type, which means that scattering off the boundary
will not change the flavor of the incoming particle. We will refer to
this type of boundary scattering problem as the diagonal case. A
natural question one would ask then is if there are more integrable
boundary conditions for these models with $O(N)$ symmetry, or if the
diagonal case exhaust all possibilities.

The purpose of this paper is to study the possible integrable boundary
conditions for models with $O(N)$ global symmetry. We find
nondiagonal integrable boundary conditions for the $O(N)$ nonlinear
$\sigma$ (nl$\sigma$) model. These are genuinely
nondiagonal boundary conditions in the sense that they can not be
put into diagonal form by a redefinition (global rotation) of the fields.
These conclusions should carry out to other models, like the $SO(N)$
Gross-Neveu and principal chiral models.

The structure of this paper is as follows. In the next section we
review some of the generalities of boundary integrable field theories
which are relevant for our discussion. We also review the
Goldschmidt-Witten argument, and use it to establish the integrability
of the $O(N)$ nonlinear $\sigma$ model. In section 3 we look at a very
simple example, to illustrate some of the issues that are relevant in
our analysis. In section 4 we introduce the new integrable boundary
conditions for the nonlinear $\sigma$ model, which are the main
concern of this paper.  In section 5 we compute the nondiagonal
reflection matrices associated to these boundary conditions by solving
the boundary Yang-Baxter equation, in particular we look at the $O(2)$
case, and present a solution for the $O(2N)$ case to which we do not
know what are the corresponding boundary conditions.  Finally in
section 6 we present our conclusions and some possible directions for
further work.

\section{Integrable Boundary Field Theory}

Let us consider a bulk integrable field theory defined by an action
\be
S= \int_{-\infty}^{+\infty} dx_1 \int_{-\infty}^{+\infty} dx_0 
\; {\cal{L}}_B(\phi,\partial \phi),
\ee
where ${\cal{L}}_B(\phi,\partial \phi)$ is the bulk Lagrangian, which
depends locally on the field $\phi$ and its partial derivatives
\footnote{Here the field $\phi$ is symbolic and may denote a
collection of fields, bosonic or fermionic fields and so on.}. We
say that it may be defined by a local action because there are
several bulk integrable models that are not, such as perturbed
conformal field theories.

As is well known, in two-dimensions integrability has several
important consequences: in a multi-particle scattering process there 
is no particle production and the set on in
and out momenta is the same, the $S$-matrix factorizes into a product of
two-body $S$-matrices, and the two-body $S$-matrix satisfies the
Yang-Baxter equation, besides the usual requirements of unitarity and
crossing-symmetry. These conditions, together with the knowledge of 
the symmetries of the $S$-matrix, often allow one to determine
the two-body $S$-matrix exactly, up to CDD factors.

When we consider this theory on the half-line, the introduction of a 
boundary may easily break the conservation of
bulk charges, as it is seen in the simple case of linear momentum. 
Therefore we are not assured that the resulting model will be integrable, and we
must specify boundary conditions. These boundary conditions correspond to
boundary dynamics, which may be described by some local lagrangian. The
boundary version of a given bulk model is defined by the action
\be
S=\int_{-\infty}^{0} dx_1 \int_{-\infty}^{+\infty} dx_0 
\; {\cal{L}}_B(\phi,\partial \phi) + \int_{-\infty}^{+\infty} dx_0 
\;{\cal{L}}_b(\left.\phi \right|_{x=0}, \left. \dot{\phi} 
\right|_{x=0}) \,\, ,
\ee
where ${\cal{L}}_b(\left. \phi \right|_{x=0}, \left. \dot{\phi} 
\right|_{x=0})$
is the boundary action, depending only on the values of the field 
$\phi$ and its time derivatives, at $x=0$ \footnote{We use an upper dot
for time derivative at the boundary.}. The boundary Lagrangian defines the
boundary dynamics, and it's easy to see that it induces boundary 
conditions through equations of motion. For example, for a scalar boson we have
\be
\partial_1 \phi |_{x=0}-\left.\frac{\partial {\cal{L}}_b}{\partial\phi}\right|_{x=0}+
\partial_0\left.\frac{\partial{\cal{L}}_b}{\partial{\dot\phi}}\right|_{x=0}=0
\,\, .
\ee

In order to have a boundary integrable model, we have to pick a
suitable boundary action such that a combination of bulk conserved
charges is still conserved. In order to discuss boundary conserved 
charges, we briefly review the bulk case.

A local (bulk) conservation law of spin $s$ is written as
\be
\pd_-J_+^{(s+1)}=\pd_+ R_-^{(s-1)}\qquad {\rm and} \qquad 
\pd_+J_-^{(s+1)}=\pd_- R_+^{(s-1)} \ , \
\ee
and from these equations, one easily establishes the conservation
of the following charges
\be
Q_+=\int_{-\infty}^{+\infty} dx_1 \; (J_+^{(s+1)}- R_-^{(s-1)} ) 
\qquad {\rm and} \qquad
Q_-=\int_{-\infty}^{+\infty} dx_1 \; (J_-^{(s+1)}- R_+^{(s-1)} )
\ee
In proving that these charges are conserved we have to use the fact
that we can discard surface terms. When we restrict our model to the
half-line we can not do that with the surface term at $x=0$.  On the
other hand, if the following condition \cite{GZ} is satisfied
\be
J_{-}^{(s+1)} -J_{+}^{(s+1)} +
R_{-}^{(s-1)} -R_{+}^{(s-1)} \x =\frac{d}{dt}\Sigma(t) 
\label{condition}
\ee
where $\Sigma(t)$ is a {\em local} field, then
\be
{\widetilde{Q}}=\int_{-\infty}^{0} dx_1 \; 
(J_- ^{(s+1)}+J_+^{(s+1)} -R_-^{(s-1)}-R_+^{(s-1)} )-\Sigma(t)
\ee
is a conserved charge. For example, in light-cone coordinates
the energy-momentum charges are $Q_{\pm}=E \pm P$, but after the
introduction of a boundary only $Q=Q_++Q_-=2E$ is conserved, with 
$\Sigma(t)=0$.

We should look for the possible boundary conditions that we can impose
to a given model such that \ref{condition} holds. Once this is done one
should look for solutions of the boundary Yang-Baxter equation
(bYBe) which can be related to the proposed boundary conditions. We
describe the bYBe in the next subsection

\subsection{The Scattering Matrix: Generalities}

Boundary integrability is encoded in the boundary Yang-Baxter equation
(bYBe) in a similar way as bulk integrability is encoded in the
Yang-Baxter equation (YBe). We should stress that, unlike the bulk
case, we do not know what is the symmetry group at the boundary a
priori. In the bulk case the knowledge of the bulk global symmetries
greatly simplify the form of the $S$-matrix.  Unfortunately we do not
have a similar set up for the boundary case, and therefore the
analysis of the possible structure of the reflection matrix will have
to rely on general physical arguments.  Recently this issue has been
addressed in \cite{MRS} for the case of the nonlinear Schroedinger
model. We review now a few well-known concepts. For a more complete
discussion, see \cite{ZZ}

The Hilbert space of in (out) asymptotic states is spanned by the 
multi-particle states
\be
|A_{i_1}(\theta_1),A_{i_2}(\theta_2) \ldots 
A_{i_n}(\theta_n)\rangle_{in(out)}=
A_{i_1}(\theta_1)A_{i_2}(\theta_2) \ldots 
A_{i_n}(\theta_n)|0\rangle_{in(out)}
\ , \
\ee
where the $\{A_i(\theta)\}$ are the Faddeev-Zamolodchikov (FZ)
operators that create the one-particle asymptotic states, the
$\{\theta_i\}$ are the rapidities and $\theta_{i_1}>\theta_{i_2}>
\ldots >\theta_{i_n}$ for in-states and the other way around for
out-states, and $|0\rangle_{in(out)}$ is the in (out) vacuum. We
assume henceforth that the in and out vacuum are the same.

Multiparticle scattering processes factorize in a product of two-body
$S$-matrix, which is defined in terms of the FZ operators by
\be
A_i(\theta_1)A_j(\theta_2)=S_{ij}^{kl}(\theta_1-\theta_2)
A_l(\theta_2)A_k(\theta_1) \ . \ \label{twobody}
\ee
The YBe is obtained by requiring the associativity of the algebra 
defined by
\ref{twobody}

In defining the reflection matrix we have to change the Hilbert space.
Boundary scattering processes are defined in terms of the boundary
state $|0\rangle_B=B|0\rangle$, where $B$ is the so-called
boundary-state operator \cite{GZ}. 

The Hilbert space is spanned by $
|A_{i_1}(\theta_1) \ldots
A_{i_n}(\theta_n)\rangle_B= A_{i_1}(\theta_1) \ldots
A_{i_n}(\theta_n)|0\rangle_B$, where the in (out) states are obtained
by setting $\theta_1>\ldots>\theta_n>0$
($\theta_1<\ldots<\theta_n<0$). Similarly to the bulk case,
the reflection matrix is defined by
\be
A_i(\theta)B=R_i^j(\theta)A_j(-\theta)B \ . \ \label{rmatrix}
\ee
We represent the $S$-matrix and the reflection matrix $R_i^j$ graphically
as in figure 1.
\vskip 0.5cm
\centerline{\epsffile{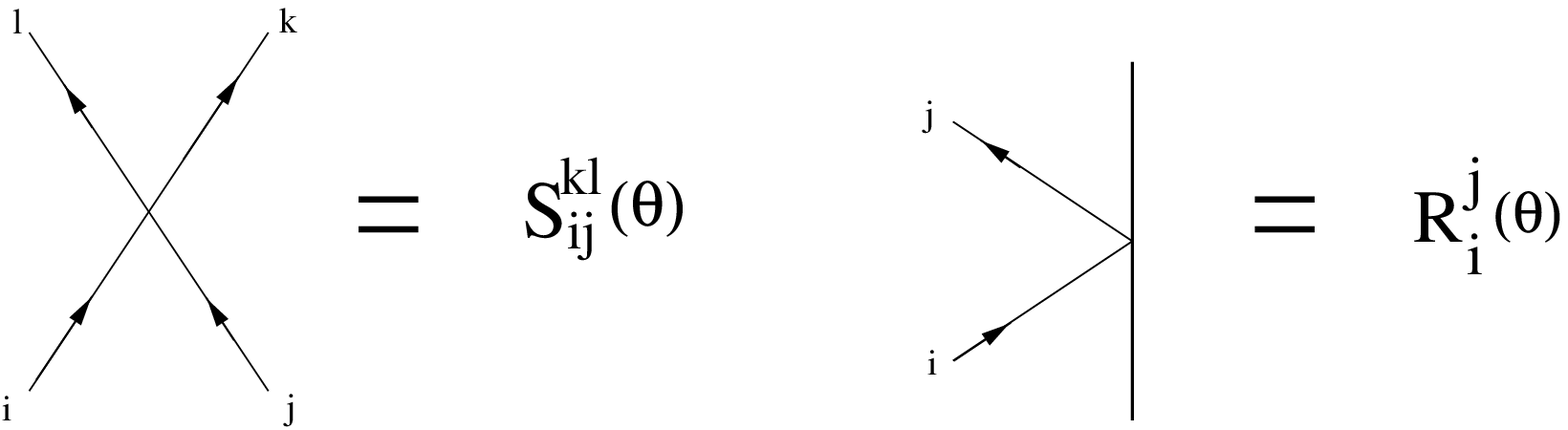}}
\vskip 0.5cm
\centerline{Fig. 1 The S-matrix and the reflection matrix}

The bYBe is obtained by requiring the compatibility of \ref{rmatrix} 
with
\ref{twobody} by looking at the process 
$|A_i(\theta_1)A_j(\theta_2)\rangle_B
\rightarrow |A_k(-\theta_2)A_l(-\theta_1)\rangle_B$. 

\vskip 0.5cm
\centerline{\epsffile{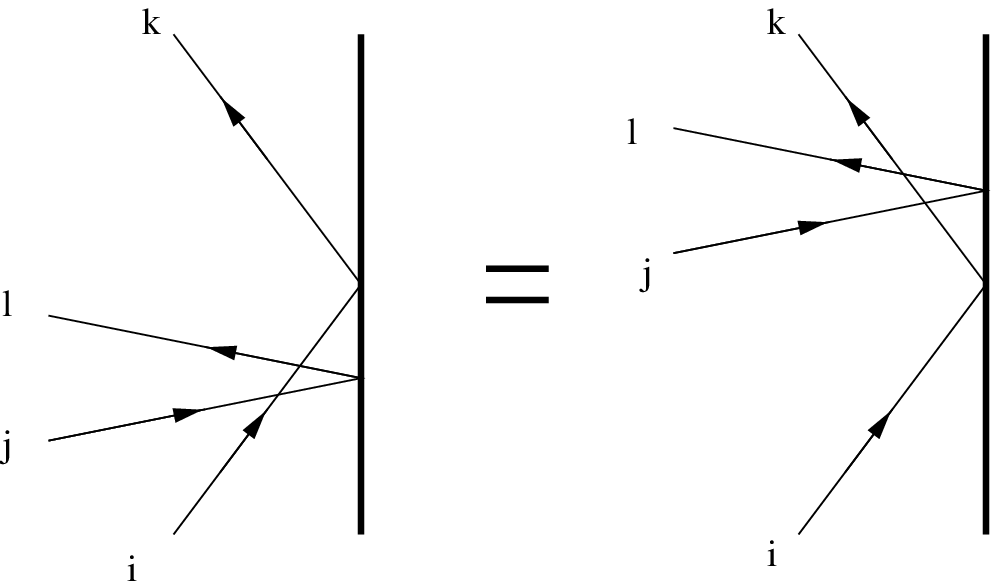}}
\vskip 0.5cm
\centerline{Fig. 2 The boundary Yang-Baxter equation.}
In terms of the two-body $S$-matrix and reflection matrix, it reads
\be
S_{ji}^{nm}(\theta)R_n^p(\theta_1)S_{mp}^{ql}(\theta_+)R_q^k(\theta_2)=
R_i^p(\theta_2)S_{jp}^{nm}(\theta_+)R_n^q(\theta_1)S_{mq}^{kl}(\theta)
\ , \ \label{bYBe}
\ee
where we introduced the variables $\theta_+=\theta_1+\theta_2$ and 
$\theta=\theta_1-\theta_2$, and sum over $m,n,p$ and $q$.

Besides the bYBe, one must impose unitary and crossing-symmetry 
conditions 
on the reflection matrix. We quote them here, and refer the reader to 
\cite{GZ}
for more details. The unitarity condition is
\be
R_i^k(\theta)R_k^j(-\theta)=\delta_i^j \label{unitarity} \ , \
\ee
and the boundary crossing symmetry is
\be
R_i^j(\frac{i\pi}{2}-\theta)=S_{kl}^{ij}(2\theta)
R_k^l(\frac{i\pi}{2}+\theta) \ , \
\ee
where we are assuming implicitly that the particles are invariant 
under charge conjugation. For the general case we refer to \cite{GZ}
and the appendix of \cite{DMM1}.

\subsection{The $O(N)$ Nonlinear Sigma Model}

In this subsection we briefly review the main features of the $O(N)$
nonlinear sigma model.

The \nls model is defined by the following Lagrangian
\be
{{\cal{L}}_{nl\sigma}}=\frac{1}{2g^2}\partial \nv \cdot \partial \nv
\ee
where $\nv$ is an $N$-dimensional vector subject to the constraint
$\nv \cdot \nv =1$. This constraint can be implemented by means of
a Lagrange multiplier $\lambda$ in the action. It is straightforward
to see that the equation of motion for $\nv$ in light-cone variables 
($x_\pm=x_0 \pm x_1$) is
\be
\partial_+\partial_- \nv = -\nv (\partial_+ \nv \cdot \partial_-\nv) \, \, . 
\ee
This will be used in establishing the integrability of the boundary conditions
of section 4.

The main features of the \nls model is that it is classically 
conformal, asymptotically free, and displays dynamical mass-generation.

\subsection{The Goldschmidt-Witten Argument}

We can use the fact that the \nls model is classically conformally
invariant in order to show that it is indeed an integrable model at
the quantum level. Starting with the classical conformal symmetry and
exploring the structure of the possible anomalies that appear after
quantization, Goldschmidt and Witten \cite{GW}, and earlier Polyakov
\cite{P}, have shown that some conserved currents
are not spoiled by quantization, that is, there {\em are} anomalies, but
their structure is such that one can rewrite the quantum corrections
to the conservation law as a total derivative, allowing us to write
down non-trivial conserved charges. This argument can be used to
establish the integrability of the Gross-Neveu model \cite{W} and of the
principal chiral model \cite{GW}. In this subsection we review the
Goldschimdt-Witten argument and write down the conserved currents that
will be of use to establish the integrability of the new boundary
conditions for the \nls model.

The Goldschmidt-Witten (GW) argument starts by observing that the 
trace of the energy-momentum of a classically conformal-invariant
theory vanishes, that is $T_{+-}=T_{-+}=0$. This implies that 
the conservation of energy-momentum reads
\be
\partial_{-} T_{++}=0 \qquad {\rm{and}} \qquad \partial_{+} T_{--}=0
\ee
which, on its turn has as a consequence that for any integer $n$
\be
\partial_{-} (T_{++})^n=0 \qquad {\rm{and}} \qquad \partial_{+} 
(T_{--})^n=0 \ . \
\label{tower}
\ee
Therefore, we generate towers of classically conserved currents.

After we quantize the theory the equations \ref{tower} do not make
sense any more, since we are considering products of operators at the
same point, and these operators should be redefined. In general these
equations will be spoiled by anomalies. The right hand side of
\ref{tower} will be nonzero, but there are still several constraints
we should impose on the possible terms that are generated: whatever
comes due to anomalies has to have the correct dimension, Lorentz
weight \footnote{If under a Lorentz transformation parametrized by
rapidity $\alpha$, a quantity $\phi\rightarrow e^{p\alpha}\phi$, we
say that $\phi$ has Lorentz weight $p$.} , and group theoretic
properties as the left-hand side. By analyzing all the possible terms
that may appear on the right-hand side of \ref{tower}, with $n=2$, GW
showed that they can all be written as total derivatives, which
implies that we have a nontrivial conserved current. 

For the \nls model they showed that the possible anomalies can be 
written
as
\be
\pd_{+}(T_{--})^2=c_1 \;\pd_+ (\pd_{-}^2\nv \cdot \pd_{-}^2\nv)+
c_2\;\pd_- (\pd_+\nv \cdot \pd_-\nv \; \pd_-\nv \cdot \pd_-\nv)+
c_3\;\pd_-(\pd_-^3\nv \cdot \pd_+\nv) \ , \ \label{chargenls}
\ee
with some constants $\{c_i\}$ and an analogous equation for 
$\pd_+(T_{--})^2$.

In \cite{MDM} we used these charges to establish the integrability of the 
aforementioned diagonal boundary conditions. We will see that we can 
use them to establish the integrability of nondiagonal boundary conditions too.

\subsection{The Exact $S$-matrix}

Once the \nls model has been established to be integrable we can proceed and compute its
two-body $S$-matrix. This has been done in \cite{ZZ2}, and 
we quote it here for further reference. Since the bulk is $O(N)$ invariant, 
the $S$-matrix has to be of the form
\be
S_{ij}^{kl}(\theta)=\sigma_1(\theta)\delta_{ij}\delta^{kl}+
\sigma_2(\theta)\delta_i^k\delta_j^l+
\sigma_3(\theta)\delta_i^l\delta_j^k
\ee
where the $\sigma_i(\theta)$ are determined by the solving the 
YBe \cite{ZZ2}, unitarity, and crossing-symmetry. In \cite{ZZ2} a large-$N$
expansion check has been performed to show that this is indeed the exact 
$S$-matrix of the $O(N)$ \nls model. The $\sigma_i(\theta)$ are given by
\bea
&&\sigma_1(\theta)=-\frac{i\lambda}{i\pi-\theta}\sigma_2(\theta) 
\qquad \ , \ \qquad
\sigma_3(\theta)=-\frac{i\lambda}{\theta}\sigma_2(\theta)
\qquad {\rm{and}} \qquad 
\nonumber \\
&&\sigma_2(\theta)=
\frac{
\Gamma(\frac{1}{2}+\frac{\lambda}{2\pi}+\frac{i\theta}{2\pi})
\Gamma(1+\frac{i\theta}{2\pi})
\Gamma(\frac{1}{2}-\frac{i\theta}{2\pi})
\Gamma(\frac{\lambda}{2\pi}-\frac{i\theta}{2\pi})
}
{
\Gamma(\frac{1}{2}+\frac{\lambda}{2\pi}-\frac{i\theta}{2\pi})
\Gamma(-\frac{i\theta}{2\pi})
\Gamma(\frac{1}{2}+\frac{i\theta}{2\pi})
\Gamma(1+\frac{\lambda}{2\pi}+\frac{i\theta}{2\pi})
} \ . \
\eea
where $\lambda=2\pi/(N-2)$.
We will also need the isoscalar $S$-matrix element 
$\sigma_I(\theta)=N\sigma_1(\theta)+\sigma_2(\theta)+\sigma_3(\theta)$,
which can be written as
\be
\sigma_I(\theta)=-\frac{(i\pi+\theta)(i\lambda+\theta)}{(i\pi-\theta)\theta}
\sigma_2(\theta) \ . \
\ee
This will be useful when we solve the boundary crossing unitarity 
condition in section 5.

For $N=2$ one has to be careful, since $\lambda \rightarrow \infty$ but
$\sigma_2(\theta) \rightarrow 0$. In this limit we get
\be
\sigma_1(\theta)=-\frac{2i\pi}{i\pi-\theta}f(\theta) \qquad , \qquad
\sigma_2(\theta)=0 \qquad , \qquad {\rm{and}} \qquad
\sigma_3(\theta)=-\frac{2i\pi}{\theta}f(\theta)
\ee
where the function $f(\theta)$ is given by
\be
f(\theta)=
\frac{\Gamma(1+\frac{i\theta}{2\pi})
\Gamma(\frac{1}{2}-\frac{i\theta}{2\pi})}{\Gamma(-\frac{i\theta}{2\pi})
\Gamma(\frac{1}{2}+\frac{i\theta}{2\pi})} \ . \
\ee
The $O(2)$ \nls model is not a simple massless free boson, as the map $n_1=\cos(\sqrt{g}\phi)$ and
$n_2=\sin(\sqrt{g}\phi)$ would suggest. It is actually the sine-Gordon model at $\beta^2=8\pi$, the
sine-Gordon potential being a marginally-relevant perturbation at this point, and it
describes the Kosterlitz-Thouless point of the classical $XY$ model.

\resection{A Very Simple Example}

Before we analyze the nondiagonal boundary conditions for the \nls
model, let us look at a very simple example of nondiagonal scattering,
consisting of 2 free bosons $\phi_1$ and $\phi_2$, with equal mass
$m$, on the half-line, which captures some of the main physical
aspects for the case of the \nls model. The action we consider is
\be
S=\int_{-\infty}^{\infty} dx_0 \int_{-\infty}^{0} dx_1 \;\;
    \frac{1}{2}(\partial \phi_1)^2 +  \frac{1}{2} m_1^2 \phi_1^2 +
    \frac{1}{2}(\partial \phi_2)^2 +  \frac{1}{2} m_2^2 \phi_2^2 +
  \int_{-\infty}^{\infty} {\cal{L}}_b(\phi_1, \dot\phi_1, \phi_2, \dot 
\phi_2) \ , \
\ee
where ${\cal{L}}_b$ is the boundary action, to be chosen 
shortly.

Clearly any quadratic form at the boundary can be solved exactly, being
obviously integrable. We have consider two possibilitites \footnote{The choice
${{\cal{L}}_b}= g_1 \phi_1^2+g_2\phi_2^2$ is trivial, since there is no
coupling between the two fields. We are also disregarding possible 
boundary mass terms.}
\bea
{{\cal{L}}_b^1}= g_1 \phi_1\phi_2  \label{L1} \\
{{\cal{L}}_b^2}= g_2 \phi_1 \dot\phi_2 \label{L2}
\eea
Recall that a boundary Lagrangian ${\cal{L}}_b(\phi_i)$ induces the 
following boundary condition
\be
\left.\frac{\partial \phi_i}{\partial x} \right|_{x=0} = 
\left.\frac{\partial {\cal{L}}_b}{\partial\phi_i}\right|_{x=0} \ . \ 
\ee

Since the masses of the two particles are the same, which is required
to have nondiagonal scattering, we can define new fields
$\eta_{\pm}$ by
\be
\eta_\pm=\frac{1}{\sqrt{2}}({\phi_1 \pm \phi_2}) \ , \
\ee
which leave the bulk action with the same form but takes the boundary
action to
\be
{{\cal{L}}_b^1} =g_1 (\eta_+^2 - \eta_-^2)
\ee
and therefore we conclude that the boundary condition induced by 
\ref{L1} is
not really nondiagonal.

On the other hand, for the second boundary Lagrangian it is not
possible to find an orthogonal transformation which diagonalizes the
boundary Lagrangian.  In this sense we can say that the boundary
Lagrangian II corresponds to nondiagonal boundary scattering.

It is fairly easy to find the reflection amplitudes for \ref{L2}. The 
boundary conditions induced by \ref{L2} are (using $g$ for $g_2$)
\bea
\left. \frac{\partial \phi_1}{\partial x}\right|_{x=0}&=&
\left. \phantom{-} g \,\, \frac{\partial \phi_2}{\partial t}\right|_{x=0}\\
\left. \frac{\partial \phi_2}{\partial x}\right|_{x=0}&=&
\left. -g \,\, \frac{\partial \phi_1}{\partial t}\right|_{x=0}
\eea
By using the mode expansion for the fields,
\be
\phi_i(x,t)=\int d\theta 
(A_i(\theta)e^{(im\cosh(\theta)t-im\sinh(\theta)x)}+
              A_i^{\dagger}(\theta) 
e^{(-im\cosh(\theta)t+im\sinh(\theta)x)})
\ee
where the $A_i(\theta)$ are the FZ operators, 
and the definition of the reflection matrix
$R_i^j(\theta)$ given in \ref{rmatrix}, we find the scattering 
amplitudes to be
\bea
R_i^j(\theta)=
\left(\begin{array}{cc}
      
\frac{\sinh^2(\theta)-g^2\cosh^2(\theta)}{\sinh^2(\theta)+g^2\cosh^2(\theta)} & 
      \frac{g\sinh(2\theta)}{g^2\cosh^2(\theta)+\sinh^2(\theta)} 
\\[2ex]
      \frac{-g\sinh(2\theta)}{g^2\cosh^2(\theta)+\sinh^2(\theta)}   & 
      
\frac{\sinh^2(\theta)-g^2\cosh^2(\theta)}{\sinh^2(\theta)+g^2\cosh^2(\theta)}
      \end{array}\right).
\eea
Notice that, as expected, the diagonal amplitudes are the same, and
the off-diagonal ones have opposite signs.  Moreover if we expand the
elements of the reflection matrix in powers of the coupling constant
$g$, the diagonal elements are even functions of $g$, whereas the
off-diagonal ones are odd functions of $g$. This can be understood
by noticing that each flavor change at the boundary is 
accompanied by a factor of $g$, and so in order to have diagonal 
(nondiagonal)
boundary scattering we need an even (odd) number of $g$'s.
We will use some of the intuition from this very
simple model in the study of the nondiagonal boundary conditions for 
the
\nls model.

\section{Boundary Conditions for the \nls Model}

As we have seen in section 3 there are essentially two possibilities we 
could try for the boundary conditions in the \nls model. 
One of them, \ref{L1}, is not even integrable in this case. 
We are left then with the boundary conditions induced by
a boundary lagrangian of the form \ref{L2},
\be
S_{nl\sigma}=S_B + 
\int_{-\infty}^{+\infty} dx_0\frac{1}{2} M_{ij}n_i \dot{n_j} \ , \ \label{action}
\ee
where we have to have $M_{ij}=-M_{ji}$, and the indices $i$ and $j$ run through a
subset of $\{1,2,\ldots, N\}$. For the remaining indices, which correspond to particles
scattering diagonally, we choose Dirichlet boundary conditions
\be
\partial_0 n_j|_{x=0}=0 \ . \ \label{bcnls1}
\ee
The reson for this choice will become clear in the next section.
From \ref{action} we get the following boundary conditions for the
non-diagonally scattering fields
\be
\partial_1\,n_i|_{x=0}=M_{ij}\partial_0\,n_j|_{x=0} \ . \ \label{bcnls2}
\ee
We have to check that the condition \ref{condition} is satisfied for
these boundary conditions, that is
\bea
&&
(\partial_-\nv \cdot \partial_-\nv)^2-(\partial_+\nv \cdot \partial_+\nv)^2+
c_1(\partial_+^2\nv \cdot \partial_+^2\nv-
\partial_-^2\nv \cdot \partial_-^2\nv)+\nonumber \\
&&
+c_2(\partial_+\nv \cdot \partial_-\nv \,\, \partial_+\nv \cdot \partial_+\nv-
\partial_+\nv \cdot \partial_-\nv \,\, \partial_-\nv \cdot \partial_-\nv)+
\nonumber \\
&&\left.
+c_3(\partial_+^3\nv \cdot \partial_-\nv-\partial_-^3\nv \cdot \partial_+\nv)
\right|_{x=0}=
\frac{d}{dt}\Sigma(t)
\eea
for some local field $\Sigma(t)$.
We discuss this condition term by term. The first term can be simplified to
\be
(\partial_-\nv \cdot \partial_-\nv)^2-(\partial_+\nv \cdot \partial_+\nv)^2=
8\,\,\nv_0\cdot\nv_1\,\,(\nv_0\cdot\nv_0+\nv_1\cdot\nv_1) \ ,
\ee
where the subscripts $0$ and $1$ denote time and space derivatives, 
respectively. The ``$c_1$'' term can be simplified to
\be
\partial_+^2\nv \cdot \partial_+^2\nv-\partial_-^2\nv \cdot \partial_-^2\nv=16\,\,\nv_{00}\cdot\nv_{01}-
8\,\,\nv_0 \cdot \nv_1 \,\,(\nv_0\cdot\nv_0-\nv_1\cdot\nv_1) \ ,
\ee
the ``$c_2$'' term to
\be
\partial_+\nv \cdot \partial_-\nv \,\, \partial_+\nv \cdot \partial_+\nv-
\partial_+\nv \cdot \partial_-\nv \,\, \partial_-\nv \cdot \partial_-\nv=4\,\,\nv_0\cdot\nv_1\,\,
(\nv_0\cdot\nv_0-\nv_1\cdot\nv_1)
\ee
and the ``$c_3$'' term to
\bea
\partial_+^3\nv \cdot \partial_-\nv-\partial_-^3\nv \cdot \partial_+\nv=
8\, \, \partial_0(\nv_{01}\cdot\nv_0-\nv_{00}\cdot\nv_1)-2\, \, \nv_0\cdot\nv_1 \,\,(\nv_0\cdot\nv_0-\nv_1\cdot\nv_1)
\eea
We see from these that we have to have $\nv_0\cdot\nv_1=0$ in order
that \ref{condition} is satisfied. Moreover, if we consider
\ref{bcnls1} and \ref{bcnls2}, we immediately see that the one ``bad''
remaining term $16\,\,\nv_{00}\cdot\nv_{01}$ vanishes.  This
establishes the integrability of \ref{bcnls1} and \ref{bcnls2}.

Since $M$ is an antisymmetric matrix, there is a real-orthogonal
matrix $O$, such that $O^t M O$ is block-diagonal, each block being an
antisymmetric $2\times2$ matrix. If $M$ has $k$ zero eigenvalues we
readily see that its structure will have $O(2) \times O(2) \times
\ldots \times O(2) \times O(k)$ symmetry, where there are $(N-k)/2$
$O(2)$'s (and therefore $N-k$ must be even).  The $O(2)$-symmetric
blocks correspond to interactions between pairs of field components at
the boundary, that is, a particle of a given type, say $i$, can
scatter diagonally or onto some other fixed particle of type $j$,
where the pairing $(i,j)$ is fixed.

The question we should try to answer now is, in which ways can we
break the boundary symmetry? In the diagonal case, the bulk $O(N)$
symmetry is broken to $O(k) \times O(N-k)$ at the boundary, with
$k=0,1,\ldots,N$. We will be looking at solutions that respect this symmetry,
and so we should consider matrices $M$ symmetric under $O(k) \times
O(N-k)$. But this implies that the only case we should consider is
$O(2) \times O(N-2)$, where the $O(2)$-symmetric part
corresponds to non-diagonal scattering, and the remaining $O(N-2)$ to
diagonal scattering, that is, the components $n_i$
satisfy
\bea
&&\partial_1\,n_1|_{x=0}=g\,\partial_0 n_2|_{x=0} \qquad , \qquad \partial_1\,n_2|_{x=0}=-g\,\partial_0 n_1|_{x=0} \nonumber \\
&&\partial_0\,n_j|_{x=0}=0 \qquad {\rm for} \qquad j=3,4,\ldots,N \ . \label{bcnls}
\eea
We have chosen the diagonally-scattering field components to satisfy Dirichlet
boundary conditions, but at this point it seems that we 
could have chosen them to satisfy Neumann. As we will see later, the solutions
of the boundary Yang-Baxter equations are such that in the ``diagonal''
limits the correct choice for the diagonally-scattering components
is indeed Dirichlet boundary conditions.

In \cite{CS} Corrigan and Sheng have established the classical
integrability of the Neumann boundary condition for the $O(N)$ \nls
model, for all $N$, and found one extra boundary condition for
$N=3$. In \cite{MDM} we showed that this special boundary condition
was integrable at the quantum level to. We will show now, that it is
actually related to the ones we propose in this paper.

The explicit form of the Corrigan-Sheng boundary condition for the $O(3)$ 
\nls model is
\be
\partial_1 \nv |_{x=0}= -k \times \partial_0 \nv |_{x=0}+ 
(\nv \cdot k\times \partial_0 \nv) \, \, \nv |_{x=0}
\qquad {\rm and} \qquad
k \cdot \partial_0 \nv|_{x=0} =0 \label{CS}
\ee
where $k$ is an arbitrary vector. Let us choose $k= g \,\, (0,0,1)$,
where $g$ is a coupling constant.  This means that $\partial_0 n_3=0$
and so $n_3$ is a constant at the boundary.  This, together with the
constraint $n_1^2+n_2^2+n_3^2=1$, imply $n_1^2+n_2^2= {\rm constant}$
at $x=0$. The Corrigan-Sheng boundary condition reduces to
\bea
&&\partial_1 n_1|_{x=0} = \phantom{-} {\tilde g} \,\, \partial_0 n_2 |_{x=0}
\qquad \ , \ \qquad
\partial_1 n_2 |_{x=0} = - {\tilde g} \,\, \partial_0 n_1 |_{x=0} \nonumber \\
&&\partial_0 n_3 |_{x=0} = 0  \ . \
\eea
This is the same boundary condition we are considering 
in the $O(3)$ \nls model case.

\section{Reflection Matrices}

In this section we will compute the reflection matrices associated to
the boundary condition identified previously. We will concentrate on
the \nls model, since the ones for the GN and PCM models, can be found
by attaching suitable CDD factors. Since we are interested in the
integrable flows for these boundary conditions, we will revisit the
diagonal case before.

\subsection{The Diagonal Case}

In \cite{DMM2} the reflection matrices associated to the diagonal
boundary conditions found in \cite{MDM} were computed (see also
\cite{G}). One question one may ask is if there are integrable {\em
diagonal} flows, that is, if it is possible to go from a configuration
with, say, $M$ components satisfying Neumann and the remaining $N-M$
Dirichlet boundary conditions, to some other other configuration with
$M'$ Neumann and $N-M'$ Dirichlet boundary conditions.  We will see
that this is not possible.

Let us suppose that there are two flavors $i$ and $j$ that scatter 
diagonally.
Consider the scattering process $|A_i(\theta_1)A_j(\theta_2)\rangle 
\rightarrow |A_i(\theta_2)A_j(\theta_1)\rangle$. Let's call the 
reflection 
amplitude for the $i$ particle $R_1(\theta)$, and for the $j$ particle 
$R_2(\theta)$
The bYBe for this process is, therefore
\bea
&&\phantom{=}R_1(\theta_1)R_2(\theta_2)\sigma_2(\theta_+)\sigma_3(\theta_-)+
R_2(\theta_2)R_2(\theta_1)\sigma_3(\theta_+)\sigma_2(\theta_-)= 
\nonumber \\
&&=R_1(\theta_1)R_1(\theta_2)\sigma_2(\theta_-)\sigma_3(\theta_+)+
R_1(\theta_2)R_2(\theta_1)\sigma_3(\theta_-)\sigma_2(\theta_+) \ . \
\eea
Dividing by 
$R_2(\theta_1)R_2(\theta_2)\sigma_2(\theta_+)\sigma_2(\theta_-)$ 
and introducing $X(\theta)=R_1(\theta)/R_2(\theta)$, and 
$\delta_i(\theta)=\sigma_i(\theta)/\sigma_2(\theta)$, we get
\be
(X(\theta_1)-X(\theta_2))\delta_3(\theta_-)=
(X(\theta_1)X(\theta_2)-1)\delta_3(\theta_+) \ . \
\ee
By taking the limit $\theta_2\rightarrow\theta_1$, we arrive at the 
following
differential equation
\be
\frac{d}{d\theta}X(\theta)=\frac{X^2(\theta)-1}{2\theta} \ . \ 
\label{ratio}
\ee
The solutions for \ref{ratio} are $X(\theta)=1$ which means that the
amplitudes $R_1=R_2$, or
\be
X(\theta)=\frac{c-\theta}{c+\theta} \ . \ \label{x_eq}
\ee
Notice that this result is true for any pair of diagonally scattering 
flavors.
Using \ref{x_eq} we will show now that there are no integrable diagonal 
flows.

Suppose there are 3 diagonally scattering flavors, and call them $1,2$ 
and 
$3$. From \ref{x_eq} we know that
\be
\frac{R_1(\theta)}{R_2(\theta)}=\frac{c-\theta}{c+\theta} \qquad \ , \ 
\qquad
\frac{R_1(\theta)}{R_3(\theta)}=\frac{c'-\theta}{c'+\theta} \qquad 
{\rm{and}}
\qquad \frac{R_2(\theta)}{R_3(\theta)}=\frac{c''-\theta}{c''+\theta}
\ee
with some constants $c, c'$ and $c''$. But these equations are
incompatible, unless one of the ratios is $1$. This proves that we can
not have more than 2 different types of reflection amplitudes in the
diagonal case.  Moreover, as it was shown in \cite{MDM}, if we assume
that there are only 2 types of reflection amplitudes, the constant $c$
in \ref{ratio} can be fixed using the bYBe to be
$c=-i\frac{\pi}{2}\frac{N-2M}{N-2}$.

Now we can easily show that there are no diagonal integrable
flows. Since the ratios are fixed by the bYBe, we simply can not
go diagonally from a reflection matrix where there are $k$ particles
satisfying Neumann boundary condition, to one where there are $k' \neq k$.

\subsection{The Nondiagonal Case}

In this subsection we use the bYBe \ref{bYBe} in order to find the
reflection matrices corresponding to the boundary conditions
identified earlier. Using the intuition gained in section 3 we have
the following ansatz for the reflection matrix
\be
R=\left(\begin{array}{ccccc}
                    \phantom{-} A(\theta) & B(\theta) & 0 & 0 & \cdots 
\\
                    -B(\theta) & A(\theta) & 0 & 0 & \cdots \\
                    \phantom{-} 0 & 0 & R_0(\theta) & 0 & \cdots  \\
                    \phantom{-} 0 & 0 & 0 & R_0(\theta) & \cdots \\
                    \phantom{-} \vdots & \vdots &  \vdots  & \vdots & 
\ddots
        \end{array}\right) \ , \ \label{r_matrix}
\ee
where $A$, $B$ and $R_0$ are to be determined. 

Let us consider the scattering process
$|A_1(\theta_1)A_i(\theta_2)\rangle \rightarrow
|A_i(-\theta_1)A_1(-\theta_2)\rangle$, where $i$ is any of the
diagonally scattering particles, that is $i>2$ The bYBe for this
process is
\bea
&&
\phantom{+}R_0(\theta_1)R_0(\theta_2)\sigma_3(\theta_+)\sigma_2(\theta)+
R_0(\theta_2)A(\theta_1)\sigma_2(\theta_+)\sigma_3(\theta)=
R_0(\theta_1)R_0(\theta_2)\sigma_3(\theta)\sigma_2(\theta_+)+ \nonumber 
\\
&&+A(\theta_1)A(\theta_2)\sigma_2(\theta)\sigma_3(\theta_+)- 
B(\theta_1)B(\theta_2)\sigma_2(\theta)\sigma_3(\theta_+) \ . \ 
\label{block}
\eea
Dividing this equation by $R_0(\theta_1)R_0(\theta_2)
\sigma_2(\theta_+)\sigma_2(\theta)$, and defining 
$X(\theta)=A(\theta)/R_0(\theta)$ and 
$Y(\theta)=B(\theta)/R_0(\theta)$, 
we get
\be
(X(\theta_1)-X(\theta_2))\delta_3(\theta)=
(X(\theta_1)X(\theta_2)-1)\delta_3(\theta_+)-
Y(\theta_1)Y(\theta_2)\delta_3(\theta_+) \ , \ 
\ee
and by taking the limit $\theta_1\rightarrow\theta_2$, we obtain the 
following
differential equation
\be
\frac{d}{d\theta}X(\theta)=\frac{X^2(\theta)-Y^2(\theta)-1}{2\theta} \ 
. \
\label{ratio1}
\ee
Notice that if $Y(\theta)=0$ this equation reduces to the one for the 
diagonal
case \ref{ratio}. Since we have two unknown functions now, we need one
more equation. This is accomplished by the bYBe for 
$|A_1(\theta_1)A_i(\theta_2)\rangle\rightarrow
|A_i(-\theta_1)A_2(-\theta_2)\rangle$, which reads
\bea
&&\phantom{+}R_0(\theta_2)B(\theta_1)\sigma_2(\theta_+)\sigma_3(\theta)=
R_0(\theta_1)B(\theta_2)\sigma_2(\theta_+)\sigma_3(\theta)+ \nonumber 
\\
&&+A(\theta_1)B(\theta_2)\sigma_2(\theta)\sigma_3(\theta_+)+
B(\theta_1)A(\theta_2)\sigma_2(\theta)\sigma_3(\theta_+) \ . \ 
\label{block2}
\eea
After dividing by $R_0(\theta_1)R_0(\theta_2)\sigma_2(\theta)
\sigma_2(\theta_+)$, and taking the limit 
$\theta_1\rightarrow\theta_2$, we get
\be
\frac{d}{d\theta}Y(\theta)=\frac{X(\theta)Y(\theta)}{\theta} \ . \ 
\label{ratio2}
\ee
We can easily solve equations \ref{ratio1} and \ref{ratio2} by introducing
$Z_{\pm}(\theta)=X(\theta)\pm iY(\theta)$, which satisfy
\be
\frac{d}{d\theta}Z_{\pm}(\theta)=\frac{Z_{\pm}^2(\theta)-1}{2\theta} \ 
. \
\ee
This is the same equation we obtained earlier for the diagonal case 
\ref{ratio}. The solution for $X(\theta)$ and $Y(\theta)$ are, 
therefore
\be
X(\theta)=\frac{1}{2}\left(\frac{c-\theta}{c+\theta}+
\frac{c'-\theta}{c'+\theta}\right) \qquad {\rm{and}} \qquad
Y(\theta)=\frac{1}{2i}\left(\frac{c-\theta}{c+\theta}-
\frac{c'-\theta}{c'+\theta}\right) \label{xy}
\ee
Notice that the purely diagonal case corresponds to $Y(\theta)=0$,
that is $c=c'$, and we recover \ref{x_eq}. One aspect that may look
puzzling is the fact that, at this point, we have two constants to be
fixed, $c$ and $c'$, but only one coupling constant $g$.

In order to find a further constraint involving $c$ and $c'$, consider
the scattering process $|A_1(\theta_1)A_1(\theta_2)\rangle \rightarrow
|A_1(-\theta_1)A_2(-\theta_2)\rangle$. The bYBe reads
\bea
&&A(\theta_1)B(\theta_2)(\sigma_2(\theta_+)\sigma_3(\theta)+
\sigma_2(\theta)\sigma_2(\theta_+))+
A(\theta_2)B(\theta_1)(\Sigma(\theta_+)\sigma_3(\theta)-
\sigma_1(\theta_+)\sigma_2(\theta))=\nonumber \\
&&A(\theta_1)B(\theta_2)(\Sigma(\theta)\Sigma(\theta_+)+
\sigma_1(\theta)\sigma_1(\theta_+))+
A(\theta_2)B(\theta_1)(\Sigma(\theta)\sigma_3(\theta_+)-
\sigma_1(\theta)\sigma_2(\theta_+)+ \nonumber \\
&&(N-2)R(\theta_1)B(\theta_2)\sigma_1(\theta)\sigma_1(\theta_+)) 
\label{cc'}
\eea
After dividing by 
$R(\theta_1)R(\theta_2)\sigma_2(\theta)\sigma_2(\theta_+)$, we plug 
the solutions \ref{xy} into \ref{cc'} to obtain, after an  elementary but
tedious calculation, that
\be
c+c'=-i\pi\frac{N-4}{N-2} \label{c+c'}
\ee
We have checked that the expressions \ref{xy} together with \ref{c+c'}
solve all the remaining bYBe's.

There are two ways to get diagonal reflection matrices from
\ref{xy} and the constraint \ref{c+c'}. The first one is by setting $c=c'$,
and we obtain the result for the diagonal case with $2$ fields
satisfying Neumann boudary condition and the remaining $N-2$ Dirichlet
\cite{MDM}. This is the deciding factor that made us choose Dirichlet
boundary conditions for the diagonally-scattering fields in section 4.
The other way to have diagonal reflection matrices is by having both
$c$ and $c'$ go to $\infty$. In this case we have $X(\theta)=1$ and we
conclude that all field components satisfy the same boundary
condition, that is, Dirichlet.

Therefore if we write
$c=-i\pi(N-4)/(2N-4)+\xi(g)$ and $c'=-i\pi(N-4)/(2N-4)-\xi(g)$, where
$\xi(g)$ is a (unknown) function of the boundary coupling constant, we
have that, for $g=0$ ($2$ Neumann, $N-2$ Dirichlet) 
we should have $\xi(0)=0$ (and so
$X(\theta)=(-i\pi(N-4)/(2N-4)- \theta)/(-i\pi(N-4)/(2N-4)+\theta)$),
and for $g\rightarrow \infty$ (all Dirichlet), $X(\theta)=1$ and so
$\xi(\infty)=\infty$. We see, then, that there are integrable flows
between diagonal reflection matrices, and that the reflection matrices
we found have the correct behaviour, as compared to the one dictated
by the microscopic boundary Lagrangian.

All that is left to do now is to solve the boundary unitarity and
boundary crossing-symmetry conditions, which will be dealt with shortly.

Based on the ``very simple example'' studied before and the solution we
found for the case of 2 nondiagonal fields, we are tempted at constructing
other solutions of the bYBe in a similar fashion as the one just done.
As we will see now, there are no such solutions.

Consider the following ansatz for the $O(5)$ model
\be
R(\theta)=\left(\begin{array}{ccccc}
                    \phantom{-} A(\theta) & B(\theta) & 0 & 0 & 0 \\
                    -B(\theta) & A(\theta) & 0 & 0 & 0 \\
                    \phantom{-} 0 & 0 & \phantom{-}A(\theta) & B(\theta) & 0 \\
                    \phantom{-} 0 & 0 & -B(\theta) & A(\theta) & 0 \\
                    \phantom{-} 0 & 0 & 0 & 0 & R_0(\theta)
                    \end{array}\right) \ , \ \label{4fields}
\ee
This reflection matrix should correspond to the following boundary action
\be
S_b=\int_{-\infty}^{+\infty} dx_0 \frac{g}{2}(n_1\dot{n_2}-n_2\dot{n_1}+
n_3\dot{n_4}-n_4\dot{n_3})
\ee
In principle we could have two different coupling constants, one for the
fields $n_1$ and $n_2$ and another for $n_3$ and $n_4$, but we are choosing
them to be equal in order to simplify the argument. Since this case is a 
special point of the general case, the absence of solutions in this case will
establish our claim that there are no solutions with more than 2 nondiagonal
fields. We will look at the $O(5)$ case, but the argument is the same for
$O(N)$ with more than on nondiagonal block.

The bYBe for the processes $|A_1(\theta_1)A_5(\theta_2)\rangle
\rightarrow |A_5(-\theta_1)A_2(-\theta_2)\rangle$ and
$|A_1(\theta_1)A_5(\theta_2)\rangle \newline \rightarrow
|A_5(-\theta_1)A_2(-\theta_2)\rangle$ are essentially the same as in
the case for 2 nondiagonal fields only, and so the ratios $A/R_0$ and
$B/R_0$ have the same form as the ones given in \ref{xy}. If we
consider now the bYBe for the process
$|A_5(\theta_1)A_5(\theta_2)\rangle \rightarrow
|A_1(-\theta_1)A_1(-\theta_2)\rangle$ we find that there is no
solution besides the diagonal one ($c=c'$) for the constants $c$ and
$c'$, and therefore there is no solution of the bYBe consistent with
\ref{4fields}.

Before we proceed, we should mention another solution of the bYBe for
even $N$, but to which we do not know how to assign microscopic
boundary conditions.

Let us consider the following block-diagonal ansatz for the reflection matrix
\be
R(\theta)=\left(\begin{array}{ccccc}
                    \phantom{-} A(\theta) & B(\theta) & 0 & 0 & \cdots \\
                    -B(\theta) & A(\theta) & 0 & 0 & \cdots \\
                    \phantom{-} 0 & 0 & \phantom{-}A(\theta) & B(\theta) & \cdots  \\
                    \phantom{-} 0 & 0 & -B(\theta) & A(\theta) & \cdots \\
                    \phantom{-} \vdots & \vdots &  \vdots  & \vdots & 
\ddots
        \end{array}\right) \ , \ 
\ee
The bYBe for the process $|A_1(\theta_1)A_3(\theta_2)\rangle \rightarrow 
|A_3(-\theta_1)A_2(-\theta_2)\rangle$ is
\bea
A(\theta_2)B(\theta_1)\sigma_2(\theta_+)\sigma_3(\theta)&=&
A(\theta_1)B(\theta_2)\sigma_2(\theta_+)\sigma_3(\theta)+\nonumber \\
&+&(A(\theta_1)B(\theta_2)+A(\theta_2)B(\theta_1)) \,\, \sigma_2(\theta)\sigma_3(\theta_+) \ .
\eea
Dividing this equation by $\sigma_2(\theta_+)\sigma_2(\theta)A(\theta_1)A(\theta_2)$, and taking the limit
$\theta_1 \rightarrow \theta_2$, we get
\be
\frac{d}{d\theta}\left(\frac{B(\theta)}{A(\theta)}\right)=
\frac{1}{\theta}\left(\frac{B(\theta)}{A(\theta)}\right) \ ,
\ee
whose solution is
\be
\frac{B(\theta)}{A(\theta)}=\alpha \,\, \theta \ ,
\ee
where $\alpha$ is a constant. We have checked that this alone solves
all the remaining bYBe's, and therefore $\alpha$ is left as a
free-parameter. All we need to do is to fix $A(\theta)$ using
unitarity and boundary crossing symmetry. Since their solution is
quite straightforward, we will not quote it here. The only diagonal reflection matrix
that can be obtained from this solution is by taking  
$\alpha=0$, which corresponds to all
components satisfying Neumann boundary conditions. If $\alpha
\rightarrow \infty$ we have a block diagonal solution, each block
being antisymmetric. Note that one must be careful in taking
$\alpha \rightarrow \infty$, since unitarity and crossing symmetry
will give a prefactor with an overall dependence on $\alpha$, which
precisely cancels the $\alpha$ dependence of $B(\theta)$ in this limit.

\subsection{The $O(2)$ Case}

In the limit $N\rightarrow 2$ we can solve the bYBe completely. In
this subsection we will not write down the bYBe's explicitly, since
they can be derived quite easily, and the $O(2)$ case is not our main interest. 
We will describe which equations one needs to solve to fix the
structure of the reflection matrix completely, in any case.

Initially consider the general form for the reflection matrix
\be
R(\theta)=\left(\begin{array}{cc}
                    f(\theta) & w(\theta) \\
                    h(\theta) & g(\theta)
        \end{array}\right) \ . \ 
\ee
The bYBe for $|A_1(\theta_1)A_1(\theta_2)\rangle \rightarrow |A_1(-\theta_1)A_1(-\theta_2)\rangle$ gives
\be
h(\theta_1)w(\theta_2)=h(\theta_2)w(\theta_1) \ . \ \label{hw}
\ee
If these amplitudes are non-zero, we have to have
\be
w(\theta)=\alpha h(\theta) \ ,
\ee
where $\alpha$ is a constant.
Next, the bYBe for $|A_1(\theta_1)A_1(\theta_2)\rangle \rightarrow |A_2(-\theta_1)A_2(-\theta_2)\rangle$
requires
\be
\frac{d}{d\theta}\left(\frac{f(\theta)}{g(\theta)}\right)=\frac{1}{2\theta}\left(\left(\frac{f(\theta)}{g(\theta)}\right)^2-1\right)
\ee
whose solution is
\be
\frac{f(\theta)}{g(\theta)}=\frac{\beta-\theta}{\beta+\theta}
\ee
for some constant $\beta$. Finally, the bYBe for  $|A_1(\theta_1)A_1(\theta_2)\rangle \rightarrow |A_2(-\theta_1)A_1(-\theta_2)\rangle$
demands
\be
g(\theta)=\gamma h(\theta) \frac{\beta+\theta}{\theta} \ . \
\ee
The reflection matrix is, then,
\be
R(\theta)=h(\theta)\left(\begin{array}{cc}
                     \gamma \frac{\beta-\theta}{\theta} & \alpha \\
                     1 & \gamma \frac{\beta+\theta}{\theta} 
        \end{array}\right) \ . \ \label{r_o2}
\ee
To obtain the solution where we would have set $h(\theta)=0$ (or
equivalently $w(\theta)=0$) in \ref{hw}, all we should do is to set
$\alpha=0$ in \ref{r_o2}. We have checked that \ref{r_o2} solves all
the remaining bYBe's. The function $h(\theta)$ may be fixed by
unitarity and boundary crossing-unitarity, but again, we are not going
to carry out this computation here.

One aspect that may look puzzling is the fact that we know that the
$O(2)$ \nls model is related to the sine-Gordon model, and \ref{r_o2}
has 3 independent parameters, unlike the boundary sine-Gordon model,
where one has only 2 independent parameters. The reason for that is
that we are not looking at the sG model at arbitrary coupling
constant, but at $\beta^2=8\pi$, and as pointed out in \cite{GZ}, at
special points for the coupling constant (like this one) there are
more solutions than the ones presented there for arbitrary $\beta$.

We now go back to the solution of unitarity and boundary-crossing
symmetry conditions for the reflection matrix \ref{r_matrix}.  The
next two subsections refer to this reflection matrix.

\subsection{Boundary Unitarity}

The boundary-unitarity condition implies that
\be
R_0(\theta)R_0(-\theta)=1 \ , \label{u1}
\ee
and
\be
A(\theta)A(-\theta)-B(\theta)B(-\theta)=1 \qquad {\rm{and}} \qquad 
A(-\theta)B(\theta)+A(\theta)B(-\theta)=0 \ . \label{u2} 
\ee
By dividing \ref{u2} by $1=R(\theta)R(-\theta)$, and using the explicit
forms \ref{xy}, we see that \ref{u2} is satisfied trivially. Therefore, 
all we have to solve from unitarity is \ref{u1}.

\subsection{Boundary Crossing Unitarity}

The boundary-crossing unitarity condition for the diagonal part of the
reflection matrix is
\be
R_0(\frac{i\pi}{2}-\theta)=((N-2)\sigma_1(2\theta)+\sigma_2(2\theta)+
\sigma_3(2\theta))R_0(\frac{i\pi}{2}+\theta)+
2\sigma_1(2\theta)A(\frac{i\pi}{2}+\theta) \label{bcu} \ . 
\ee
Using \ref{xy} and \ref{c+c'}, we can rewrite \ref{bcu} as
\be
R_0(\frac{i\pi}{2}-\theta)=R_0(\frac{i\pi}{2}+\theta)(\sigma_I(2\theta)
+2\sigma_1(2\theta)(X(\frac{i\pi}{2}+\theta)-1)) \label{bcu2} \ .
\ee
All the other boundary-crossing unitarity equations are equivalent to
this one.Using the parametreization for $c$ and $c'$ introduced after
equation \ref{c+c'}, and after some elementary, but tedious,
computation, we can cast \ref{bcu2} into a quite compact form
\be
R_0(\frac{i\pi}{2}-\theta)=R_0(\frac{i\pi}{2}+\theta)\sigma_I(2\theta)
\left[\frac{\theta+\xi-\frac{i\lambda}{2}}{\theta-\xi+\frac{i\lambda}{2}}\right]
\left[\frac{\theta-\xi-\frac{i\lambda}{2}}{\theta+\xi+\frac{i\lambda}{2}}\right] 
\label{bcu3}
\ee
Finally we can fix $R_0(\theta)$ by solving \ref{bcu3} and \ref{u1}
\bea
&&R_0(\theta)=
-\frac
{\Gamma(\frac{1}{4}+\frac{\lambda+2i\xi}{4\pi}+\frac{i\theta}{2\pi})
 \Gamma(\frac{3}{4}+\frac{\lambda+2i\xi}{4\pi}-\frac{i\theta}{2\pi})}
{\Gamma(\frac{1}{4}+\frac{\lambda+2i\xi}{4\pi}-\frac{i\theta}{2\pi})
 \Gamma(\frac{3}{4}+\frac{\lambda+2i\xi}{4\pi}+\frac{i\theta}{2\pi})}
\times \nonumber \\
&&\phantom{R_0(\theta)=-}
\frac
{\Gamma(\frac{1}{4}+\frac{\lambda-2i\xi}{4\pi}+\frac{i\theta}{2\pi})
 \Gamma(\frac{3}{4}+\frac{\lambda-2i\xi}{4\pi}-\frac{i\theta}{2\pi})}
{\Gamma(\frac{1}{4}+\frac{\lambda-2i\xi}{4\pi}-\frac{i\theta}{2\pi})
 \Gamma(\frac{3}{4}+\frac{\lambda-2i\xi}{4\pi}+\frac{i\theta}{2\pi})}
K(\theta)
\eea
where $K(\theta)$ is the solution of \ref{bcu3} and \ref{u1} without 
the $\xi$-dependent factors \cite{DMM2},
\be
K(\theta)=
-\frac{\Gamma(\frac{1}{2}+\frac{\lambda}{4\pi}-\frac{i\theta}{2\pi})\,
\Gamma(1+\frac{i\theta}{2\pi})\,
\Gamma(\frac{3}{4}+\frac{\lambda}{4\pi}+\frac{i\theta}{2\pi})\,
\Gamma(\frac{1}{4}-\frac{i\theta}{2\pi})}
{\Gamma(\frac{1}{2}+\frac{\lambda}{4\pi}+\frac{i\theta}{2\pi})\,
\Gamma(1-\frac{i\theta}{2\pi})\,
\Gamma(\frac{3}{4}+\frac{\lambda}{4\pi}-\frac{i\theta}{2\pi})\,
\Gamma(\frac{1}{4}+\frac{i\theta}{2\pi})}
\ee
This fixes the reflection matrix completely. 

This solution for the reflection matrix reduces to the appropriate
diagonal solutions in the limits $\xi \rightarrow 0$ ($2$ Neumann, $N-2$
Dirichlet) and $\xi \rightarrow \infty$ (all Dirichlet).

It would be interesting to study the analytical structure of this reflection 
amplitude, but this requires the knowledge of the explicit form of the 
function $\xi(g)$.

\section{Conclusions}

We have found new integrable boundary conditions for the $O(N)$ \nls
model which are nondiagonal and depend on one free parameter. The
integrable boundary conditions presented here break the $O(N)$
symmetry to $O(2) \times O(N-2)$. A similar conclusion has been
reached for the $SU(N)$ spin chain in \cite{AR}.  These boundary
conditions, together with the diagonal ones proposed in \cite{DMM2},
are argued to exhaust the possible integrable boundary conditions for
the $O(N)$ \nls model. We also found solutions of the bYBe which
depend on one parameter and that can be associated to these boundary
conditions, and saw that they do satisfy the appropriate limits, which
correspond to the coupling constant $g\rightarrow 0$ or $\infty$. We
found the most general solution of the bYBe for the $O(2)$ \nls model,
which is described by a three-parameter family, unlike the well-known
two-parameter solution of the boundary sine-Gordon. This is understood
by noting that the $O(2)$ \nls model corresponds to the $\beta^2=8\pi$
sine-Gordon model, and as pointed out in\cite{GZ}, at this point, and
in general for points where $\lambda=8\pi/\beta^2-1$, there are
additional solutions.

In section 4 we established the integrability of the nondiagonal
boundary conditions, by inspecting the spin-$4$ charges proposed in
\cite{GW}. There were no constraints on how many nondiagonal blocks
one can have, whereas we found that there are solutions of the bYBe
only in the case of one single nondiagonal block. One possible
explanation for that is that by inspecting higher spin charges there would be
further constraints on the possible boundary conditions, but this is still to
be checked.

Recently, MacKay and Short \cite{MS} have studied the principal chiral model
with a boundary and found an interesting relationship between their
boundary conditions and the theory of symmetric spaces. Their
solutions, though, are quite different from ours, and some work should
be done in trying to clarify their relationship.

There are several directions to be pursued now. The $S$-matrix for the
elementary excitations in the $SO(N)$ principal chiral and $O(N)$
Gross-Neveu models, are, up to CDD factors, the same as the one for
the $O(N)$ \nls model, and therefore the bYBe's for all the three
models is the same. This means that it should be possible to find
microscopic lagrangians which provide nondiagonal boundary conditions
for these other models, similar to the ones presented here, also
depending on one free parameter.

Another problem that arises naturally
is to find the exact form of the function $\xi(g)$, which is, in
general quite hard (as for example, in the case of the boundary
sine-Gordon model). A related problem is to perform a large-$N$
expansion for the reflection matrices (diagonal and nondiagonal) as a
check of their validity.  Unfortunately a framework of how to perform
boundary perturbation theory is yet to be developed. Finally, the
study of the thermodynamics of the boundary \nls model would be an
interesting, if somewhat challenging, problem.

\section*{Acknowledgments}
I would like to thank T. Becher, A. De Martino, P. Fendley,
T. Jayaraman, A. Petrov, V. Sahakian, and G. Thompson for
discussions. The hospitality of the Abdus Salam ICTP, George Thompson,
and SISSA are gratefully acknowledged. This work is in part supported by 
the NSF.

\end{document}